\documentclass[twocolumn,twocolappendix,trackchanges]{aastex631}
\pdfoutput=1

\shorttitle{CO Line Profiles in Supernova Remnants}
\shortauthors{Fuda, Tram and Reach 2021}
\graphicspath{{./}{figures/}}
\usepackage{amsmath}
\usepackage{graphicx}
\usepackage{siunitx}
\usepackage{multirow}
\usepackage{booktabs}
\usepackage{vcell}
\usepackage{subfigure}
\submitjournal{ApJ}
\accepted{10-Jan-2023}
\begin{document}

\def\bfc{\bf}

\title{Modeling CO Line Profiles in Shocks of W28 and IC443}

\author[0000-0002-6372-8395]{Nguyen Fuda}
\affiliation{Lunar and Planetary Laboratory, University of Arizona, 1640 E. University Boulevard, Tucson, AZ 85721, USA}
\email{fudanguyen@email.arizona.edu}

\author[0000-0002-6488-8227]{Le Ngoc Tram}
\affiliation{Max Planck Institute for Radio Astronomy, 53121 Bonn, Germany}
\email{nle@mpifr-bonn.mpg.de}


\author[0000-0001-8362-4094]{William T. Reach}
\affil{Universities Space Research Association, MS 232-11, NASA Ames Research Center, Moffett Field, CA 94035, USA}
\email{wreach@sofia.usra.edu}

\begin{abstract}

Molecular emission arising from the interactions of supernova remnant (SNR) shock waves and molecular clouds provide a tool for studying the dispersion and compression that might kick-start star formation as well as understanding cosmic ray enhancement in SNRs. Purely-rotational CO emission created by magneto-hydrodynamic shock in the SNR--Molecular cloud interaction is an effective shock tracer, particularly for slow-moving, continuous shocks into cold inner clumps of the molecular cloud. In this work, we present a new theoretical radiative transfer framework for predicting the line profile of CO with the Paris-Durham 1D shock model. We generated line profile predictions for CO emission produced by slow, magnetized C-shocks into gas { of density} $\sim 10^4$ cm$^{-3}$ {with shock speeds of 35 and 50$\,\rm km\,s^{-1}$}. The numerical framework to reproduce the CO line profile utilizes the Large Velocity Gradient (LVG) approximation and the omission of optically-thick plane-parallel slabs. With this framework, we generated predictions for various CO spectroscopic observations up to $J=16$ in SNRs W28 and IC443, obtained with SOFIA, IRAM-30m, APEX, and KPNO. We found that CO line profile prediction offers constraints on the shock velocity and preshock density {independent of  the absolute line brightness and}
requiring fewer CO lines than diagnostics using an rotational excitation diagram.
\end{abstract}


\section{Introduction} \label{sec:Intro}

Believed to be the site of cosmic ray enhancements, supernova remnant (SNR) shocks provide a possible explanation for the excess cosmic ray photons and increased ionization rates in observations of SNR-molecular clouds interactions \citep{drury_power_2017}. Understanding cosmic ray origin and enhancements in SNRs provides an understanding of the SNR energy budget and production mechanisms of MeV-TeV cosmic rays. Analysis have been carried out extensively for SNR such as W28, with ROSAT X-ray observations \citep{rho_ROSATObservationW28_1996}, or the gamma-ray observations with CANGAROO \citep{rowell_ObservationsSupernovaRemnant_2000}. A more recent study of \citep{phan_constraining_2020} also shown the cosmic ray spectrum in the Northeastern W28 is best explained by cosmic ray protons from hundreds MeV to tens of TeV. The interaction of SNR and molecular clouds thus are top laboratories for constraining the cosmic ray spectrum across the Galactic plane.

Shocks into molecular clouds are laboratories for tracing the exchange of energies in the interstellar medium. Supernova shock waves carve cavities in the ISM and excite the molecular and atomic gas to higher temperatures. The gases act as coolants for the shocks, providing tracers of physical and chemical properties within molecular clouds. Observing the interactions in SNRs also provides clues about the supernova progenitor, particularly with the innermost layers of the supernova ejecta and the inferred mass and nucleosynthesis profiles \citep{temim_ProbingInnermostEjecta_2019}.

{The Paris-Durham MHD-shock model \citep{gusdorf_sio_2008,lesaffre_low-velocity_2013, tram_h2_2018,flower_interpreting_2015,godard_models_2019} is the integrated toolkit for studying shocks. As the Paris-Durham shock unambiguously predicts the excitation diagram (or Boltzmann diagram, relation of column density of excited levels vs. excitation energy), previous studying shocks relied on observing multiple transitions of the same molecule in a region \citep[see e.g.][]{gusdorf_probing_2012,2019A&A...628A.113L} and comparing them with either single \cite[e.g.,][]{van_der_tak_computer_2007} or multiple gas temperature and density \citep[e.g.,][]{2002MNRAS.332..985L}.}

{Tracing shocks with molecular spectroscopy is of great importance given their roles in diffusing molecular clouds and also compressing parts of the clouds and potentially triggering star formation. The slow, continuous shocks (C-type; \citealt{draine_theory_1993,draine_magnetohydrodynamic_1983}) have been shown to be particularly important for explaining the existence of OH-masers due to their non-dissociative nature: the resulting thick, warm, broad shock-front allowed for efficient cooling compared to the thin layers of dissociative jump shocks (J-type) \citep{wardle_SupernovaRemnantOH_2002}. Such interactions are observationally constrained by molecules. The Paris-Durham shock code can explicitly predict the velocity-resolved profile (or line profile) for an optically thin line such as H$_{2}$ \cite[e.g.,][]{reach_supernova_2019} or optically thick line such as SiO \citep[see e.g.,][]{gusdorf_sio_2008, gusdorf_sio_2008-1}.}

{Carbon monoxide (\(\rm^{12}C^{16}O \)) is among the most abundant molecules in the ISM, second only to diatomic hydrogen (H\(_2\)). 
In contrast with H$_2$, CO can be excited in low-temperature (even under $10\,$K) and low-density gas (even under $10^3\,$cm$^{-3}$), including the range of temperatures and densities expected for dense cores in molecular clouds (10--100$\,$K and $10^5\,$cm$^{-3}$, respectively). The pure-rotational transitions (\(\rm\triangle J=1\)) of the CO molecule cover far-infrared to sub-millimeter wavelengths. These properties complement H$_{2}$ as an obsrevational tracer, with CO emission lines  capabile to trace the cold, dense gas (see Table \ref{tab:co_rotational_transitions}) that could provide information insight into star-forming cores \citep{burton_ic_1987,van_dishoeck_submillimeter_1993}.}

{In this work, we expand the theoretical line profile modeling to CO, devise a custom numerical radiative transfer integration scheme from shock models, and apply it to new CO observations in IC443 and W28 - two pristine sites for studying the interaction between shocks and molecular clouds.}

This paper is organized as follows: In Section \ref{sec:NatureShock}, we review observational constraints in SNR W 28 and IC 443 and present new spectroscopic observations with SOFIA in the W28-BML4 region (broad molecular emission region). In Section \ref{sec:MHDShockModels}, we discuss the theory of magneto-hydrodynamics (MHD) shock and the setup of a-priori Paris-Durham shock models. In Section \ref{sec:RadTransPredictions} we discuss the associated theoretical radiative transfer framework and the relevant numerical scheme. In Section \ref{sec:ResultLineProfile} we applied the framework to generate line profile prediction of various CO observations to constrain shock properties, such as preshock density and shock velocity.  
\begin{figure*}[p]
\centering
\includegraphics[width=.7\textwidth]{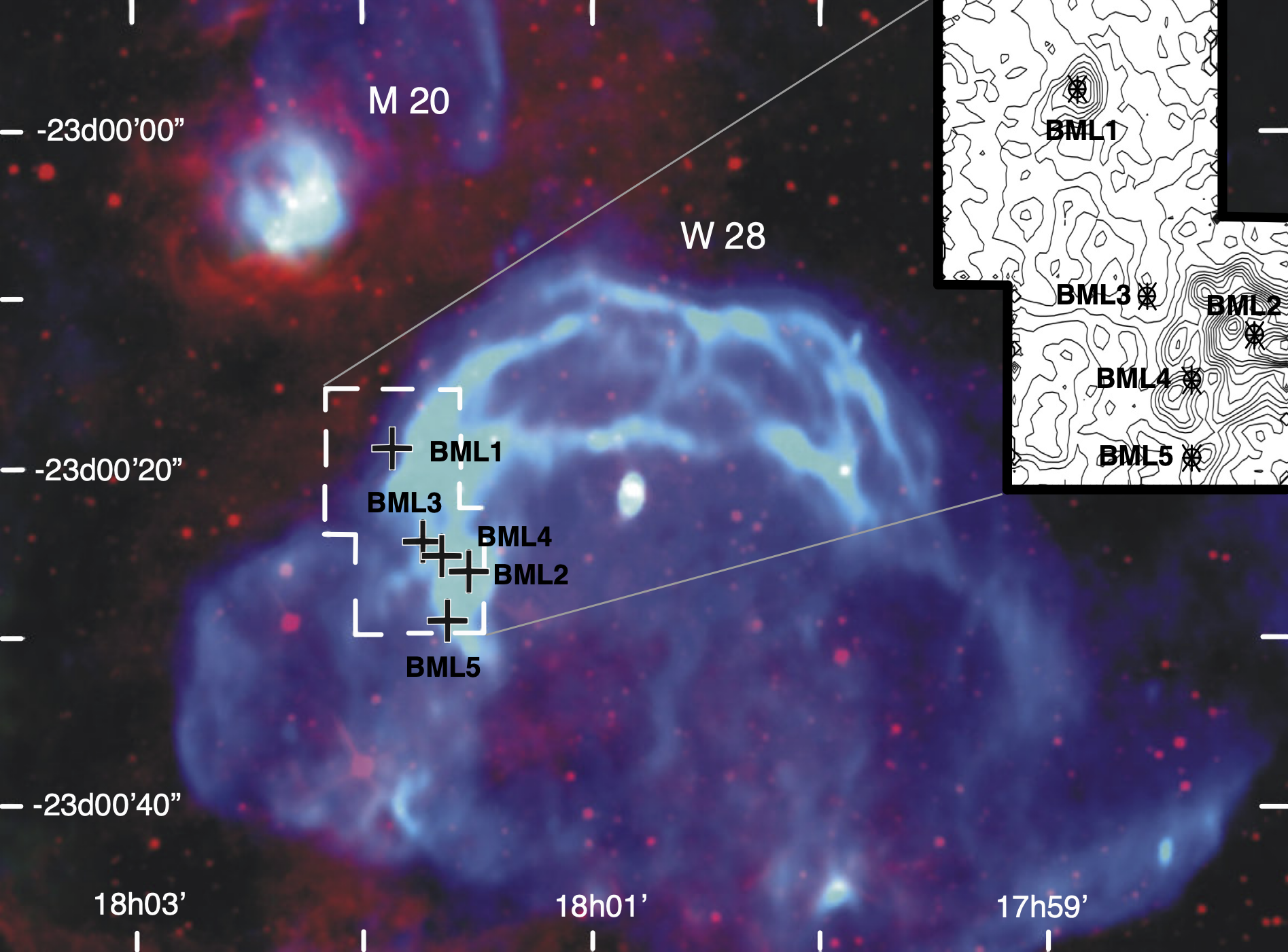}
\caption{False color image (RGB with band VLA(P) 90 cm, MSX 8 $\mu$m, VLA(L) 20 cm) of the SNR W 28 and its northern shocked ring \citep{brogan_oh_2004}. The northeastern, dashed polygon indicates the region of the overlaid CO 2-1 map \citep{reach_shocked_2005} with the 12-meter ARO Telescope on Kitt Peak (KP12). Positions of 5 broad molecular lines (BML) regions are shown. 
\label{fig:w28_nrao_image}}

\includegraphics[width=.6\textwidth]{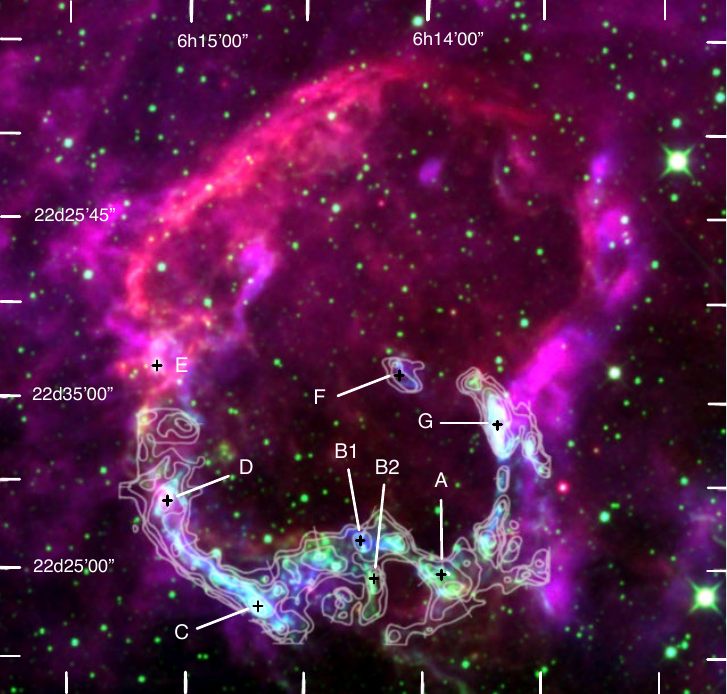}
\caption{WISE IR false-color image of SNR IC 443 (band W4 22 $\mu$m, W2 4.6 $\mu$m, W3 12 $\mu$m) and its entire shock rings. Overlaid are the contour map of the 2.1 \(u\)m H$_2$ emission from \citep{burton_ic_1987} for the \(\omega\)-shaped southern ridge, with shocked clumps positions indicated.
\label{fig:ic443_wise_image}}
\end{figure*}

\begin{deluxetable}{ccccc}
\tablecaption{The critical density of CO for pure-rotational transitions.
\label{tab:co_rotational_transitions}}
\tablewidth{\textwidth}
\tablehead{
\colhead{$J_{\rm up \rightarrow low}$} & 
\colhead{Energy} & \colhead{$\nu_{\rm up-low} $}	& \colhead{$n_{\rm crit}$(10K)} & \colhead{$n_{\rm crit}$(50K)} \\
\nocolhead{} & \colhead{(K)} & \colhead{(GHz)} & \colhead{($\rm cm^{-3}$)}  & \colhead{($\rm cm^{-3}$)} \\
}
\startdata
2 $\rightarrow$ 1 & 16.60 & 230.54 & \num{9.7e3} & \num{1.2e4} \\
3 $\rightarrow$ 2 & 33.19 & 345.79 & \num{3.2e2} & \num{3.8e2} \\
5 $\rightarrow$ 4 & 82.97 & 576.27 & \num{1.8e5} & \num{1.7e5} \\
9 $\rightarrow$ 8 & 248.88 & 1036.91 & \num{1.0e6} & \num{8.8e5} \\
11 $\rightarrow$ 10 & 364.97 & 1267.01 & \num{1.6e6} & \num{1.5e6} \\
13 $\rightarrow$ 12 & 503.13 & 1496.92 & \num{2.2e6} & \num{2.2e6} \\
16 $\rightarrow$ 15 & 751.72 & 1841.35 & \num{3.5e6} & \num{3.5e6} 
\enddata
\tablecomments{Calculated from the LAMDA database as a ratio of radiative coefficient and collisional coefficient (n\(_{\rm crit}= A_{ij}/C_{ij}\), where H$_2$ is the main collision partner \(n_{\rm H_2}=\num{1e4}\) cm\(^{-3}\)). Zero-indexing is used. Pure-rotational lines (\(\Delta J=\pm1\)) of CO  trace both cool and dense gas as well as hot gas in shocks, indicated by the excitation temperature and critical density for collisional excitation. High-J CO lines specifically indicate regions of denser, colder CO concentration.}
\end{deluxetable}

\section{Nature of shocks in W28 and IC 443} \label{sec:NatureShock}


\subsection{Shocks in W28}\label{subsec:obs_w28}

 W28 is a mixed-morphology SNR, among the brightest  in radio and $\gamma$-ray emission
  \citep{giuliani_agile_2010,green19}. The region where shocks interact with the dense interstellar clouds is complex, with evidence of shocks into gas with a range of densities \citep{reach_shocked_2005}. Its mixed morphology is seen by the contrast between the non-thermal radio shell structures versus thermal X-ray emission from the interior and requires an age greater than $10^3$ yr \citep{rhopetri98} and more likely of 
  order $4\times 10^4$ yr \citep{giuliani_agile_2010}. Multi-wavelength studies from sub-mm \citep{dubner_high-resolution_2000,vaupre_cosmic_2014} to X-ray \citep{zhou_xmm-newton_2014} and $\gamma$-ray have progressively reveal different aspects of the supernova-molecular cloud interaction and acceleration of cosmic rays\citep{cui18}.
\cite{pannuti_ctio_2017} studied optical lines and X-ray sources in the interior of W28 and found evidence for oxygen-rich ejecta, suggesting that its progenitor star was massive. Furthermore, they pointed out that X-ray sources near the center of the SNR are unlikely to be remnants of the progenitor. 

Strong evidences of the interaction with SNR blast wave with dense gas in the northern ridge of W 28 have been presented by mm detection of 1720 MHz OH maser in W 28 \citep{claussen_polarization_1997}.
Using mm and near-infrared detection of rotational/ro-vibrational  lines of CO, CS and HCO$^+$, H$_2$ and atomic lines, \citet{reach_shocked_2005} gave an estimate of n(\(\text{H}_{2})\sim 10^{3}-10^5 \text{ cm}^{-3}\) for the density of observed regions in W28, based on the observation of the broad molecular lines (BML) regions, and they inferred that a large fraction of mass of the molecular clouds lies in a relatively compact volume. The observed SiO linewidth (\(\sim 21 \text{ km s}^{-1}\)) and brightness also provided a lower limit of shock velocity at \(V_s \geq 20 \text{ km s}^{-1}\). Furthermore, \cite{reach_shockingly_1996} calculated the line ratios of CS(2-1) / CO(2-1) \(\sim 0.04\) and CS(3-2) / CS(2-1) \(\sim 1.0\), which further constrained the excitation condition to \(T\sim100 \text{K}\) and n(\(\text{H}_{2}\)) \( \leq 10^5\text{cm}^{-3}\). 

More recent observation of various rotational CO lines with SOFIA-GREAT towards the location of maser OH(F) has revealed multiple overlapping shocked components, suggesting propagation into a complex cloud structure spanning multiple range of densities. \cite{gusdorf_probing_2012} made use of multiple rotational lines to build a shock-predictive excitation model. The excitation diagram methods in \cite{gusdorf_probing_2012} involved observing six CO lines of both $^{12}$CO and $^{13}$CO and predicting shock density using the excitation of each rotational level computed by a non-LTE radiative transfer with a single density-temperature input.
Best-fit models indicate a stationary C-shock configuration, with $n=10^4$ cm$^{-3}$ at an age of $10^4$ years. To explain the line-width, a sum of two shock velocities plus a ambient gas layer are used: V$_{\rm{Shock}} =20$ and $25$ km s$^{-1}$ with magnetic field components  $b=0.4 $ and 1 - or 40 and 100 $\mu$G, respectively. Projections effects are suggested to be the explanation for the difference in the two values. Upper estimate of post-shock density at \num{3e5} cm$^{-3}$  increases the magnetic field strength to 550 $\mu$G. The lower velocities limits are a consequence of the 1D nature of the model and the lack of direct constraints in the line profile. 

\begin{figure}[ht!]
\plotone{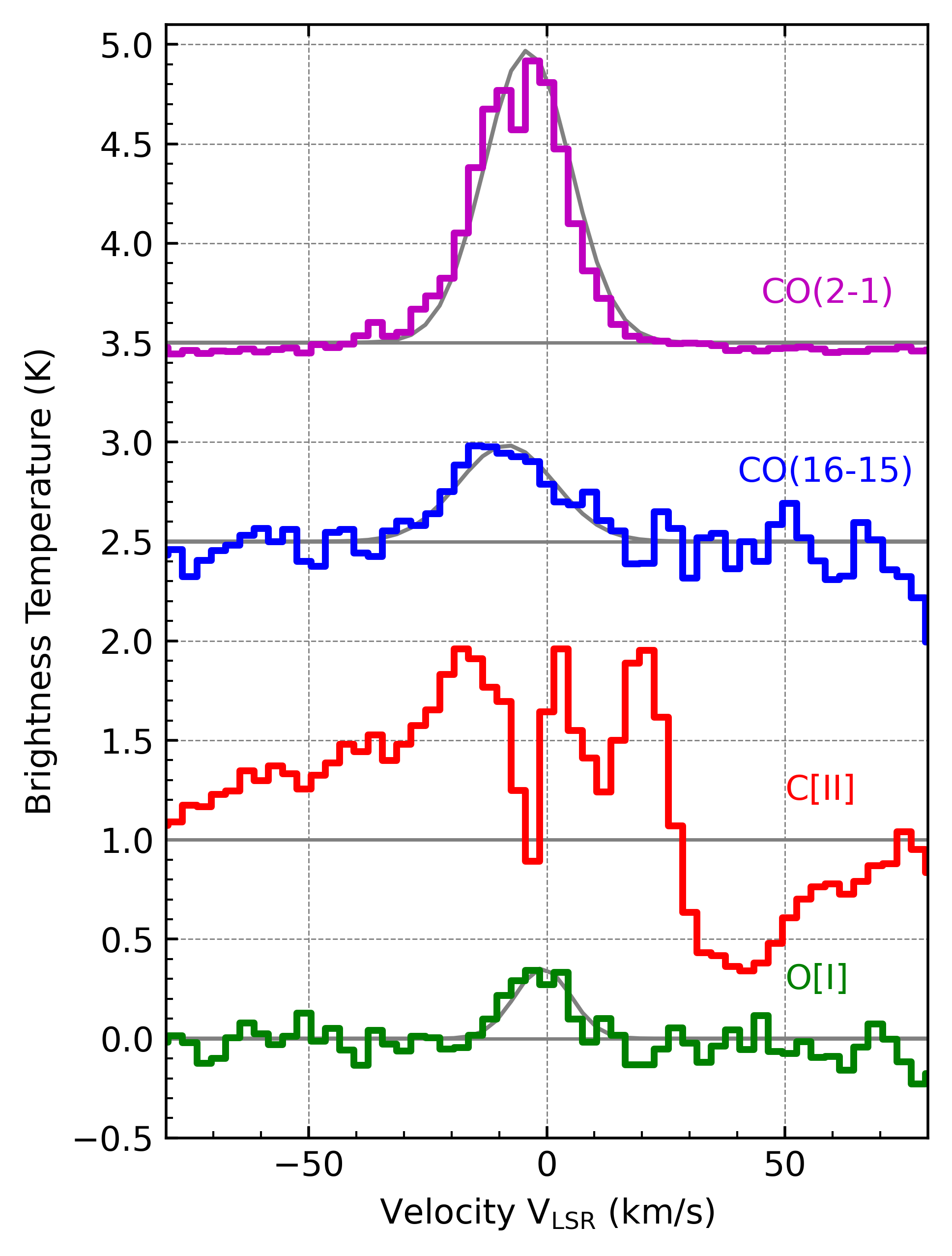}
\epsscale{1.2}
\caption{Comparison of new SOFIA observations of the BML4 region of W28 in the CO(16-15), [\ion{C}{2}], and [\ion{O}{1}] (see Table \ref{tab:sofia_observations_properties}) with the IRAM 30-m observations of CO(2-1) (see Table \ref{tab:other_observation_properties}). Thin solid lines show Gaussian fits. The [\ion{C}{2}] emission is complicated by foreground gas on this line of sight that absorbs the emission from shocked gas in W28.
\label{fig:w28_new_sofia_observation}}
\end{figure}

{ 
In this work, we combine archival observations with new SOFIA observations of W28 BML4, at J2000 coordinates 18:01:40.3 -23:25:03, the brightest region of CO(2-1) emission \citep{reach_shocked_2005}.
Observations were made using the German REceiver At Terahertz frequencies \citep[][GREAT]{great_ref} on 2018 May 23, from 40,000 feet altitude, as part of program 06\_0001. 
Figure \ref{fig:w28_new_sofia_observation} shows the new spectra of CO(J=16-15), $\rm O_I$ ($^{3}\text{P}_{J=2-1})$ at 63 $\mu$m, and $\rm C_{II}$ ($^{2}\text{P}_{J=3/2-1/2}$) at 158 $\mu$m. 
The properties of spectral lines from the new SOFIA observations are summarized in Table \ref{tab:sofia_observations_properties}, and those from archival observations summarized in Table \ref{tab:other_observation_properties}.
}

All of these lines showcase similarly broad linewidth, as seen with low-J CO observations in the same region. 
Additionally, the unmatched central velocity peaks of high-J CO and low-J CO (10 and 5 km s$^{-1}$ respectively) in this region likely suggest different origins of shock, as previously observed in clump F of W28 by \cite{gusdorf_probing_2012}. It could be seen by comparing the centers of CO(16-15) observation with the CO(2-1) observation from \cite{reach_shocked_2005} (see Figure \ref{fig:w28_new_sofia_observation}). The line ratio of \(\langle \mathrm{C_{ II}} \rangle/\langle  \mathrm{O_{ I}} \rangle\sim 5\) provides a constraint on the range of {\it a priori} shock fits.

\begin{deluxetable}{cccccc}
\tablecaption{Properties of spectral lines from new SOFIA observations of W28 BML4.
\label{tab:sofia_observations_properties}.}
\tablewidth{\textwidth}
\tablehead{
\colhead{Line}  & \colhead{$V_{\rm cen}$} & \colhead{FWHM} & \colhead{$T$} & \colhead{$\int TdV$} & \colhead{Beam}\\
&  \colhead{(km s$^{-1}$)}  & \colhead{(km s$^{-1}$)} & \colhead{(K)} & \colhead{(K km~s$^{-1}$)} & \colhead{($"$)}\\
}
\startdata
  {\rm CO(16-15)}  & 0.1 & 15 & 0.4 & 2.2 & 15\\
  {\rm [\ion{C}{2}]}  & -20 & 60 & 0.9 & 16 & 14\\
  {\rm [\ion{O}{1}]}   & -7 & 20 & 0.45 & 9.1 & 6\\
\enddata
\end{deluxetable}

\begin{deluxetable}{ccccchcc}
\tablewidth{\textwidth}
\tablecaption{Properties of spectroscopic observations of W 28 (BML4, OH maser F) and IC 443 (G, H) used for modeling. 
\label{tab:other_observation_properties}}
\tablehead{
\nocolhead{} & \colhead{Line}  & \colhead{T} & \colhead{$V_{\rm cen}$} & \colhead{FWHM} & \nocolhead{$\int TdV$} 
&\colhead{Beam}
&\colhead{Ref.\tablenotemark{a}}\\
\colhead{~} & \colhead{~} & \colhead{(K)} & \colhead{(km~s$^{-1}$)}  & \colhead{(km~s$^{-1}$)} & \nocolhead{(K km s$^{-1}$)} 
&\colhead{($''$)}
& \nocolhead{}
}
\startdata
\multicolumn{7}{c}{\textbf{W 28}} \\ \cmidrule{1-8}
\multirow{2}{*}{\centering BML4} &  CO(3-2) & 1.8 &  -6.9 & 12 & 40        &  18  & \multirow{2}{*}{1}\\
                                  &  {\rm CO(2-1)} & 45 &  +9 & 26 & 1210  &  11   &\\ 
\cmidrule{1-8}
 \multirow{2}{*}{\centering OH-F} &  CO(3-2) & 13.5 &  -0.8 & 27 & 323 & 18 & \multirow{2}{*}{2} \\
                                &  CO(11-10) & 2.2 &  -5 & 13    & ??  & 12 &  \\
\hline                                
\multicolumn{7}{c}{\textbf{IC 443}} \\ 
\cmidrule{1-8}
 \cmidrule{1-8}
 \multirow{2}{*}{\centering G} & CO(2-1) & 30 &  -8.0 & 15.0 & 504   & 31 & \multirow{2}{*}{\centering 3}\\
                                 & CO(3-2) & 39 &  -8.0 & 15.0 & 504 & 21 &\\
\cmidrule{1-8}
  H1 &  {\rm CO(2-1)} & 1.3 & -3 & 15 & 25  & 27 & \multirow{2}{*}{\centering 4}\\
  H3 &  {\rm CO(2-1)} & 1.9 & -3 & 17 & 25 & 27 & \\
\enddata
\tablenotetext{a}{1: \cite{reach_shocked_2005}, 2: \cite{gusdorf_probing_2012}, 3:\cite{van_dishoeck_submillimeter_1993}, 4: \cite{rho_near-infrared_2001}}
\end{deluxetable}

\subsection{Shocks in IC443}\label{subsec:obs_ic443}

IC 443 is another example of a supernova explosion adjacent to molecular clouds, which produces shock interactions across a range of densities. X-ray studies have revealed the { compact stellar remnant of the} progenitor of IC443 to be a neutron star \citep{olbert01,swartz15}. An extremely well-studied object, IC 443 is a textbook example of shock-molecular cloud interaction.

Its morphology is mixed, with pristine shell-like structures where shocks interact with molecular clouds in the Northeastern and Southwestern regions, all indicated by emission from in X-ray, sub-mm, and far-infrared. 
Early study by \cite{van_dishoeck_submillimeter_1993} revealed that shocks mainly occur at dense regions forming a flat ring shape in the southern portion of the SNR. 
The observations in the sub-mm are best explained by shocks into clumps of gas, each accelerated differently based on densities. There is evidence that shock into high-density gases is a potential accelerator of cosmic rays up to TeV energies, through the correlation of gamma-ray observation and shock into molecular gas. 

The preshock material likely takes a range of densities from lower density outer shells to higher density cores. Observations and theoretical modeling supports the presence of gas with preshock density \(\sim 10^3- 10^4 \rm cm^{-3}\) in the southern ridge of IC 443. Observations of H$_2$ and CO emission \citep{van_dishoeck_submillimeter_1993,burton_ic_1987,rho_near-infrared_2001, cesarsky_isocam_1999} revealed clumps of various densities within IC 443, whose preshock properties are mainly probed with excitation models.  Mid-IR 6.9 \(\rm \mu m\) observation and modeling of H$_2$ S(5) prediction of \cite{reach_supernova_2019} further strengthen a similar range of preshock density using direct H$_2$ line profile fitting. In this work, we make used of CO observations from \cite{van_dishoeck_submillimeter_1993} and \cite{rho_near-infrared_2001} in concordance with shock properties predicted with various molecular lines emission.

Age estimates of \cite{troja_xmm-newton_2008} based on X-ray analysis is at t $\sim$ 4,000 years, while \cite{chevalier_supernova_1999} optical observations put it much older at t $\sim$ 30,000 years. However, more recent efforts of theoretical modeling of 1D MHD-shock for infrared observation \citep{reach_supernova_2019} and 3D shock for X-ray observation \citep{ustamujic_modeling_2020} both supported age estimates in the range of t $\sim$ 3000--8000 years. In this study we employed a similar range of age estimates of 3000--10,000 years.

\section{MHD Shock Modeling}\label{sec:MHDShockModels}

\citet{draine_theory_1993} review interstellar shocks in the interstellar medium, in which supernova explosions are amongst the major sources.  In the rest frame of the shock, its structure can be divided into three regions: the pre-shock region, the radiative zone, and the post-shock region. The preshock material is heated and ionized by upstream (`precursor') radiation and cosmic ray. As it reaches the shock transition, deceleration in flow velocity occurs (in the shock rest frame). In the radiative zone, kinetic energy is transferred into thermal energy, and entropy increases strongly. The post-shock gas cools down as it moves further from the shock, where gas and dust could be irradiated by photons that propagated far downstream.

In scenarios such as the interaction of shock fronts with molecular clouds, strong magnetic fields are usually present. The Paris-Durham shock model\footnote{\url{https://ism.obspm.fr/shock.html}} \citep{gusdorf_sio_2008,flower_interpreting_2015,godard_models_2019} enables modelling for such magnetic-dominated shocks with, assuming that the B--field is transverse to the direction of propagation.

For two-fluid shocks, the magnetic fields are responsible for separating the flow into neutral flow, and ionic flows which experience magnetic drag, coupling to the magnetic fields. Despite taking up small fraction in mass, ionic flow strongly affect shock dynamics. For two-fluids shock, three types of solutions are possible: C (continuous shock), J (jump shock), and a mixed CJ (or C*), depending on whether the shock transition smoothly from super- to subsonic speed as it decelerates \citep{draine_theory_1993,draine_magnetohydrodynamic_1983}.


We generated four a-priori C-shock models to investigate the resulting line profiles. Properties of CO(2-1) line profiles for each shock model are shown in Table \ref{tab:shockmodels}. We generated a priori C-shock models with preshock density $\num{5e3} \leq$ n(H$_{\rm 2}$ $\leq \num{2e4}$ cm$^{-3}$, shock velocity 35 and 50 km s$^{-1}$, dimensionless magnetic factor b $=B \: (n_{H})^{-1/2} \sim 2.5, 2.8$ $\mu$G cm$^{-1.5}$, and shock ages estimates from 5,000-10,000 years. Line-profile calculations will require separate post-processing, which is discussed in detail in Section \ref{sec:RadTransPredictions}. 

{
Besides the key shock properties mentioned in the previous paragraph, we used the following for all models. The cosmic ray ionization rate $\zeta$ is chosen to be $2\times 10^{-15}$ s$^{-1}$ based upon elevated cosmic ray rates measured
near IC 443 \citep{indriolo10}. The initial conditions of the gas were adopted
from those of a steady-state photodissociation region model for gas density $2\times 10^3$ cm$^{-3}$, solar-circle interstellar radiation field, and low extinction ($A_{\rm V}=0.1$). These preshock conditions result in a temperature 47 K, which is higher than a nominal, quiescent molecular cloud, far from the supernova. 
The thickness of the shocked layer depends upon the temperature and chemical state of the pre-shock gas, which may not be well known. For this work, we use the high cosmic-ray ionization rate measured specifically in the molecular gas near IC 443.
}

The Paris-Durham shock models can readily predict the brightness of atomic lines, such as those of the neutral oxygen [\ion{O}{1}] and the singly-ionized carbon atom [\ion{C}{2}]. They are included as additional shock diagnostics specifically for the W28-BML4 observations, as C and O data are available. Calculation from \cite{reach_shocked_2005} and \cite{draine_theory_1993} indicates that $\rm O_{I}$ intensity of $10^{-4}, 10^{-3}, 10^{-5}$ ergs s$^{-1}$ cm$^{2}$ sr$^{-1}$ -- or equivalently, 10, 11 and 9 K for brightness temperature -- corresponds respectively to very fast shock, dissociative shock, and magneto-hydrodynamic shocks -- the latter of which produce the majority of molecular lines. 

Generally, strong C shocks with velocity \(V_{\rm Shock}<60\) km s\(^{-1}\) are primarily non-dissociative (i.e. H$_2$ survives) for a range of density from \(10^2\) -- \(10^5\) cm\(^{-3}\), since the heating rate due to ion-neutral friction remains low enough, whereas strong J-type shocks are mainly atomic \citep[refer to Figure 2 of][]{mckee_interstellar_1991}. Multi-fluid shocks are the main mechanism to accelerate molecules to high velocities without being dissociated. Specifically, C-shocks are crucial to explaining the origin of large-linewidth molecular rotation-vibration transitions observed in shocked regions of molecular clouds, and thus will be the target of our focus in this study.

\begin{deluxetable*}{clcccccccc}
\tablewidth{0pt}
\tablecaption{
Properties of a-priori C-shock models. \label{tab:shockmodels}.}
\tablehead{
\colhead{No.} & \colhead{$n$(H$_2$)} & \colhead{$V_{\rm shock}$} & \colhead{b} & \colhead{Age} & \colhead{$\langle \rm{C_{II}} / \rm{O_{I}} \rangle$} & \colhead{Pre-shock Cut} & \colhead{Post-shock Cut} & \colhead{FHWM} & \colhead{$\int TdV$} \\
\nocolhead{} & \colhead{(cm$^{-3}$)} & \colhead{(km s$^{-1}$)} & \colhead{($\mu$G cm$^{-1.5}$)} & \colhead{(yr)} & \nocolhead{} & \colhead{($\num{1e14}$ cm)} & \colhead{($\num{1e15}$ cm)} & \colhead{(km s$^{-1}$)} &\colhead{(K km s$^{-1}$)} 
}
\startdata
    (1) & 20,000 & 50 & 2.8 & 5,000 & 0.1 & 3 - 6 & 2 & 19.40 & 7.71  \\ 
    \cmidrule{2-10}
    (2) & \multirow{2}{*}{10,000} & 50 & 2.8 & 5,000 & 7 & 5 - 8 & 2 & 22.07 & 6.5 \\
    (3) &                         & 35 & 2.5 & 5,000 & 6 & 8 - 20 & 5 & 16.05 & 10.38 \\ 
    \cmidrule{2-10}
    (4) & 5,000 & 35 & 2.8 & 5,000 & 50 & 9 - 20 & 10 & 14.72 & 0.84 \\
\enddata
\tablecomments{
Densities and shock velocity are chosen based on prior works \citep{reach_supernova_2019,van_dishoeck_submillimeter_1993} and serve as baselines for demonstration of the theoretical model, as opposed to precision measurement. The integrated intensity of CO emission from each shock model is included as a sanity check with observation, on the basis that scale factors from 0.1 to 2 is reasonable.
Also included are the properties of the prediction: the LVG integration range and the FWHM. 
}

\end{deluxetable*}
%
\section{Predictions of Velocity Distribution of CO in W28 and IC443}\label{sec:RadTransPredictions}
\subsection{Large Velocity Gradient Approximation}\label{subsec:LVG}

In this work, we made use of the \emph{Large Velocity Gradient (LVG)} approximation introduced by \cite{sobolev_diffusion_1957}. In an expanding shell such as in SNR, if the Doppler shift due to velocity gradient between gas cells is larger than the local thermal line width, a photon may find itself `escaping' from absorption \citep{surdej_contribution_1977,hummer_sobolev_1985,sobolev_diffusion_1957}. 

To solve the radiative transfer equation, we firstly find the coefficients for emission and absorption which depend on the level population of each line. The level population \(n\) (\(\rm cm^{-3}\)) is derived from the `on the spot' assumption that incoming and outgoing transitions
from each energy level are balanced while the gas is at the same spatial location.
Moreover, with the use of the LVG approximation, the probability that photons might escape from absorption is represented by \(\beta_{ij}\) for transition  \(i\rightarrow j\). 
The population balance equation is determined by:
\begin{eqnarray}
&\beta_{i+1,i} n_{i+1} A_{i+1,i} + (n_{i+1}B_{i+1,i} -n_{i}B_{i,i+1}) \beta_{i+1,i} I_c 
\nonumber \\
-&[\beta_{i,i-1} n_i A_{i,i-1} + (n_{i}B_{i,i-1} -n_{i-1}B_{i-1,i}) \beta_{i+1,i} I_c ] 
\nonumber \\
&+ \sum_{k\neq i}(n_k C_{ki}-n_i C_{ik})=0    .
\end{eqnarray}
where $A$ and $B$ are the Einstein spontaneous and stimulated emission rate coefficients, and 
\(C\) is the collisional rate coefficient for transitions at given temperatures obtained from laboratory measurements;  we obtained the collisional and radiative transition rates and from the Leiden Atomic and Molecular Database (LAMDA\footnote{\url{https://home.strw.leidenuniv.nl/~moldata/}}).
 \(I_{c}\) is the Planck function \((2h\nu^3/c^2)/(h\nu/k T -1))\) evaluated at the background temperature \(T_{\rm CMB}=2.73 \:K\), which dominates at microwave frequencies \citep[][see A.3]{gusdorf_sio_2008}.

An approximation for the escape probability is given by \cite{neufeld_radiative_1993} as $ \beta_{ij} = [1-\exp(-|3\tau_{\nu}|)]/|3\tau_{\nu}| $. The radiation field can be approximated by $ \bar{J}_{ij}=S_{ij}(1-\beta_{ij})+\beta_{ij} I_{c}$  where $S_{\nu(ij)}=j_{\nu}/{\alpha_\nu}$ is the source function. 

\begin{figure}[!htb]
\centering
\plotone{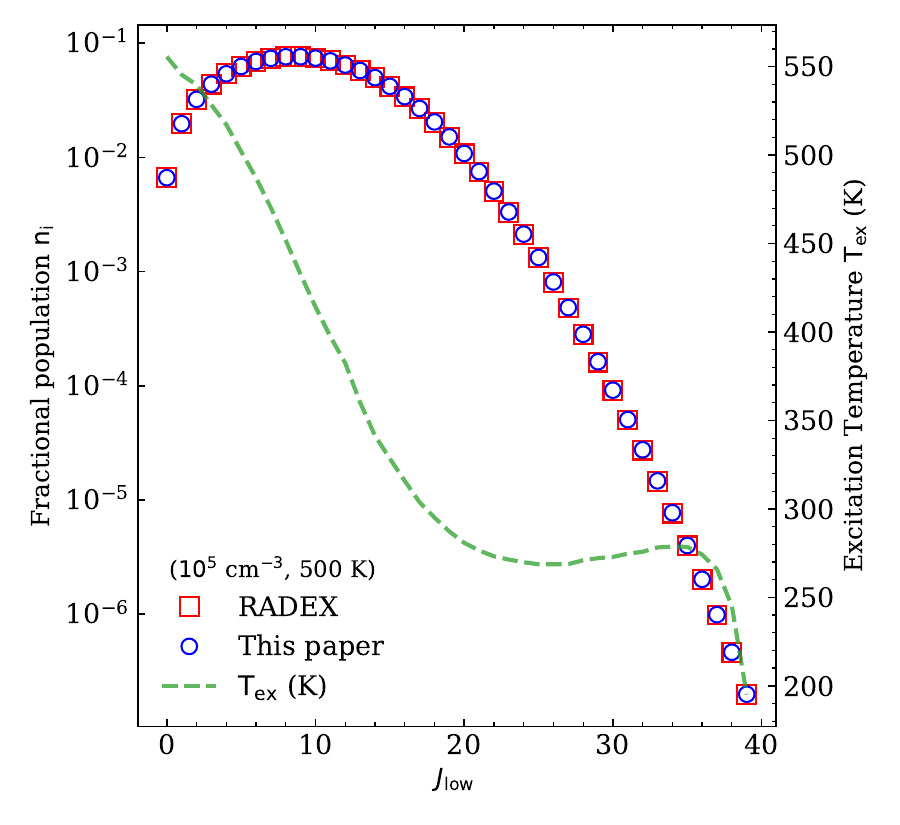}
\caption{Excitation diagram showing the fractional population of CO for various rotational levels (quantum number J=0--40) at n(H$_2$)\(\rm \sim10^{5} \: cm^{-3}\), T\(=500K\), computed by the LTE routine in this work and RADEX \citep{van_der_tak_computer_2007}. The dashed line and right-hand axis show the excitation temperature, which is significantly lower than the kinetic temperature for levels with $J>10$.
\label{fig:lte_pop}}
\end{figure}

The escape probability approximation neglects radiative energy transport from one part of the medium to the next. By combining single-point level population from the escape probability method with a simple first-order \cite{olson_short_1987} (OK87) numerical scheme (see Equation \ref{eq:ok87_o1}), 
a line profile could be modeled by finding \(I_\nu\) for each value of $v_{\rm r}$. Past work of \cite{gusdorf_sio_2008} in creating fast FORTRAN routines has provided computationally-inexpensive, grid-based calculation of the population of CO up to \((J=41)\).
In Figure \ref{fig:lte_pop}, we compared the line intensities derived in this work with the LVG approximation with the well-used RADEX code of \cite{van_der_tak_computer_2007}

\begin{figure*}[htb!]
{%
    \includegraphics[width=0.5\textwidth]{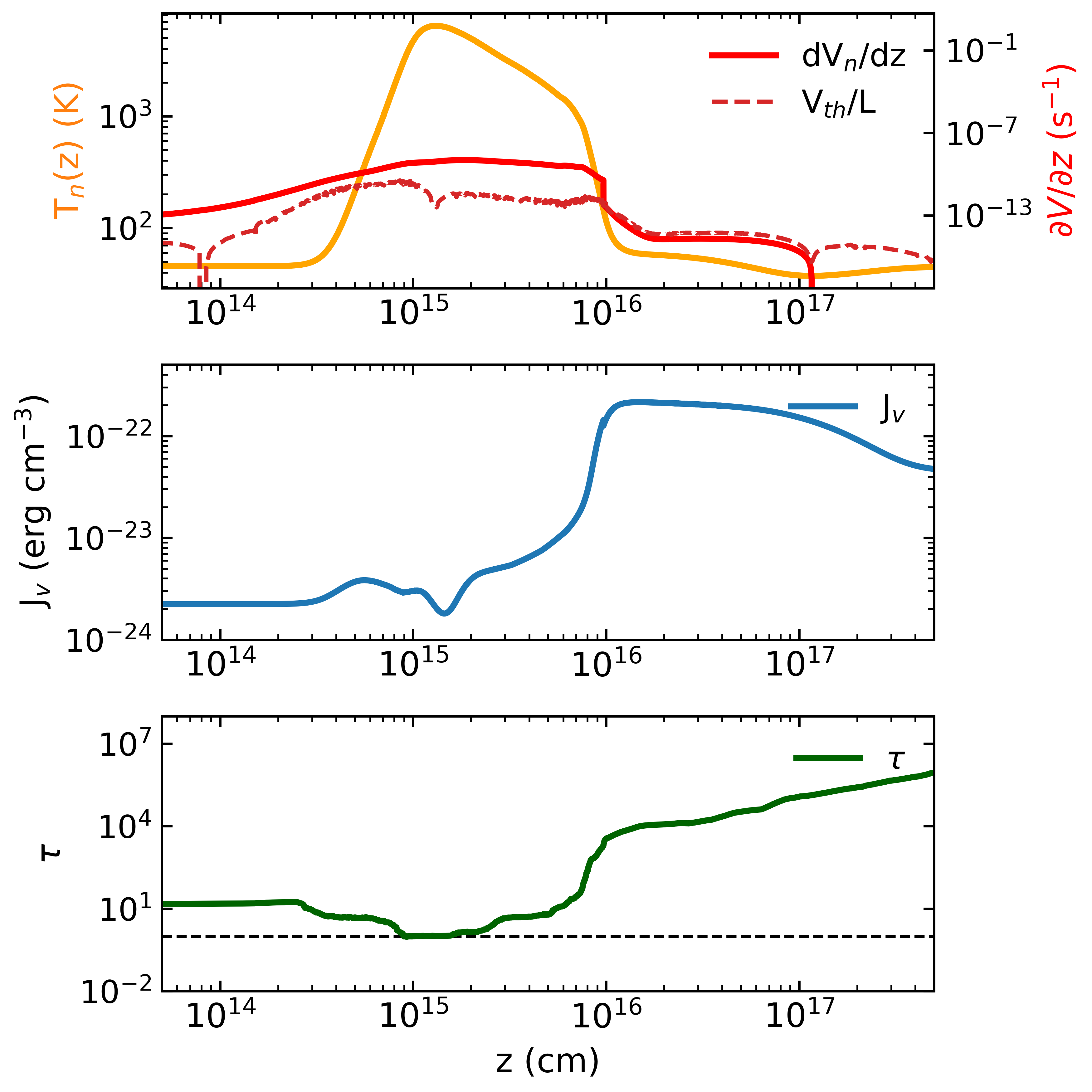}}
{%
    \includegraphics[width=0.5\textwidth]{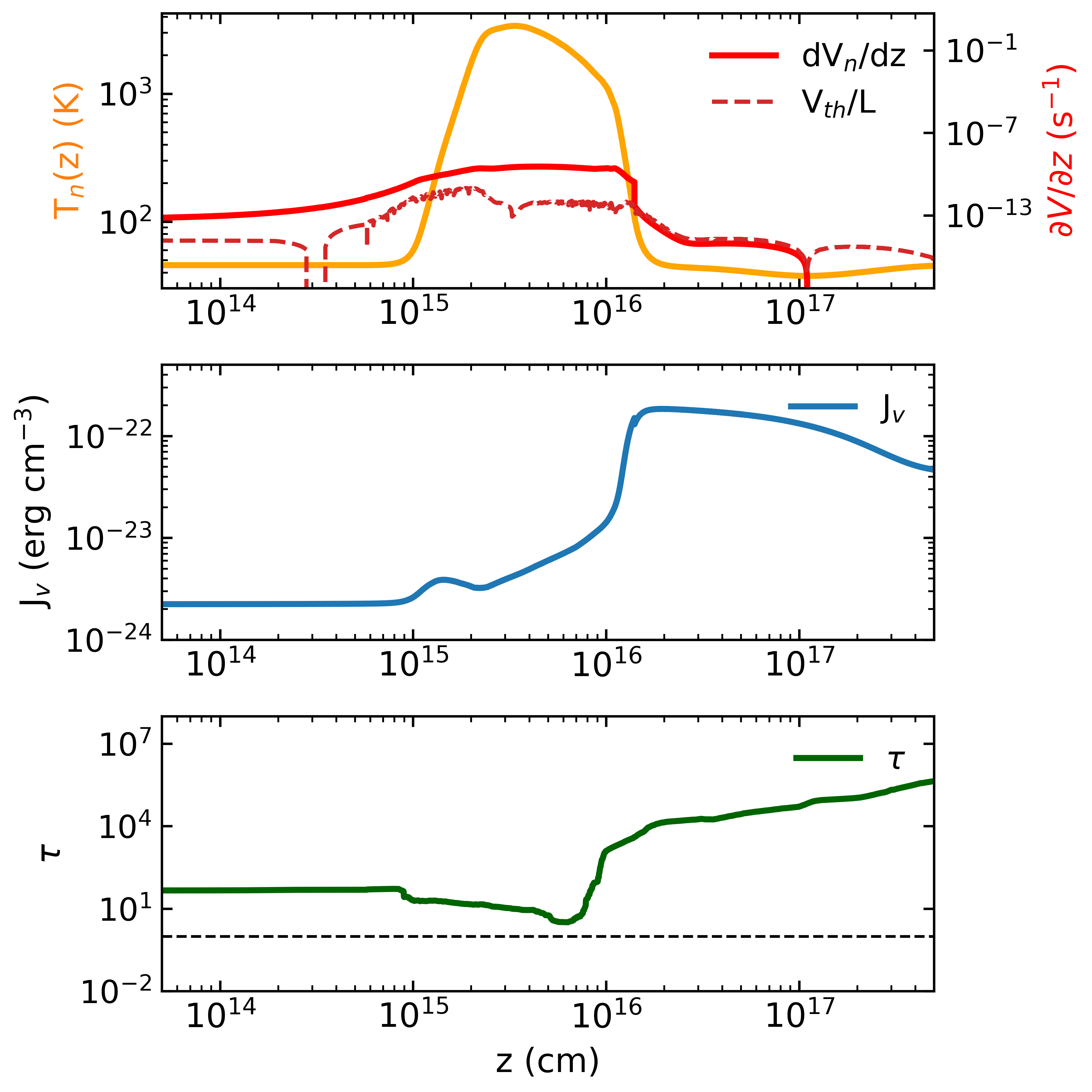}}
\caption{
Investigating the appropriate ranges for radiative transfer for C-shocks models with ({\it left column}) $ \rm V_{\rm shock}=50$ km s$^{-1}$ and ({\it right column}) $ \rm V_{\rm shock}=35$ km s$^{-1}$ into gas with density $10^4$ cm$^{-3}$.
\textit{Top row:}: LVG condition in shock. In the Paris-Durham model, the LVG-transition occur during the cooling process, after which it no longer applies.
\textit{Middle row:} The emission coefficient \(j_\nu\). 
\textit{Bottom row:} The slab optical depth. Optical-depth $\tau$ forms into two clear regimes of high-and-low magnitude around z $\sim 10^{16} $ cm for both models. \label{fig:lvg}
}
\end{figure*}


\subsection{Radiative Transfer}\label{subsec:radtrans}

Radiative transfer within gases are dominated by emission \& absorption lines. For line transfer, the emission and extinction coefficient for a transition from \(i\rightarrow j\) are given by the level population, \(n\):
\begin{equation}
j_{ij,\nu}=\frac{h \nu_{ij}}{4\pi}n_{i}A_{ij}\phi_{ij}(\nu)
\end{equation}and
\begin{equation}
\alpha_{ij,\nu}=\frac{h\nu_{ij}}{4\pi}(n_{j}B_{ji}-n_{i}B_{ij})\phi_{ij}(\nu)
\end{equation}
where \(A_{ij}, B_{ij}\) is the Einstein coefficients for spontaneous and stimulate emission. They are related such that \(A_{ij} = \frac{2h\nu^{3}_{ij}}{c^{2}}B_{ij}\) (\(\rm s^{-1}\)) and \(g_{i}B_{ij} = g_{j}B_{ji}\), where \(g\) is the statistical weight of a given level.

The photon that is emitted or absorbed may not be exactly at the same frequency \(\nu\) due to many broadening effects. The line broadening function \(\phi^*(\nu)\),  the ``probability'' that a photon from a given transition is observed at frequency \(\nu\), is such that \(\int_{0}^{\infty}\phi^*(\nu)d\nu=1\). In shock, the broadening of emission lines is due to effects such as the Doppler shift due to various velocity components in the shock region and the inclination with respect to the line of sight. In this work, all the velocity components sums up as an observer-frame velocity \(v_{\rm obs}\), such that the broadening function \(\phi(\nu)\) is given by
\begin{equation}
\phi\left(v_{r}\right)=\frac{\lambda_{ij}}{\sqrt{2 \pi} \sigma} \exp \left[-\frac{\left(v_{\mathrm{obs}} \cos \theta-v_{r}\right)^{2}}{2 \sigma^{2}}\right]
\end{equation}where \(v_{\rm obs}=v_{\rm gas}(z) - v_{\rm shock} - v_{\rm preshock}\) (the velocities are: gas parcel velocity at position \(z\); shock velocity in shock frame; and the ambient velocity in the pre-shock environment, respectively). The angle $\theta$ is from the line of sight to the shock normal. 
The shock-frame gas velocity \(v_{\rm gas}\) at position \(z\) is given by the neutral gas evolution in the shock model, whereas the shock and preshock velocity are fitting parameters. The line-width due to thermal velocity is given by 3-d Planck thermal dispersion \(\sigma=\sqrt{ 8kT /\pi \mu m_{\rm H} }\) where $\mu$ is the molecular mass in amu.

Finally, then, the rate of change of the specific intensity \(I_{\nu}\) \((\rm erg \: cm^{-1}\: s^{-1} \:sr^{-1} \: Hz^{-1})\) along the line-of-sight \(s\) can be found. Radiative transfer in a medium is described by an ordinary differential equation:
\begin{equation}\frac{dI_{\nu}}{ds} = j_{\nu}-\alpha_{\nu}I_{\nu}\end{equation}

For optically thin cases, the absorption coefficient can be discarded. The intensity then is found simply by integrating the emission coefficient over the distance \(z\), projected as \(z=s \cdot cos(\theta)\).
\begin{equation}
I_\nu =\int _{z_f}^{z_i}\: j_{\nu} dz 
\end{equation}
More complex numerical integration methods are employed for an optically thick transition. The first-order, \(\tau\)-grid method from \citet{olson_short_1987} (denoted OK87) assumes that the source function \(S_i=j_i/\alpha_i\) stays constant within a cell, but can change from cell to cell. If we integrate along direction \(z\) normal to the parallel planes, the intensity is:
\begin{equation}
\label{eq:ok87_o1}
I_{i+1/2}=e^{-\Delta \tau_{i}} I_{i-1/2}+\left(1-e^{-\Delta \tau_{i}}\right) S_{i-1/2} 
\end{equation}
where
\begin{equation}
\Delta \tau_{i}=\left(z_{i+1/2}-z_{i-1/2}\right) \frac{\alpha_{i}}{cos(\theta)}
\end{equation}
For the second-order OK87 solution, the source function may vary linearly within the cell. In this case, the intensity is:
\begin{equation}
\label{eq:ok87_o2}
I_{i+1 / 2}=e^{-\Delta \tau_{i}} I_{i-1 / 2}+Q_{i}
\end{equation}
where
\begin{eqnarray}
    Q_{i}=\left(\frac{1-\left(1+\Delta \tau_{i}\right) e^{-\Delta \tau_{i}}}{\Delta \tau_{i}}\right) S_{i-1 / 2}
    \nonumber \\
    + \left(\frac{\Delta \tau_{i}-1+e^{-\Delta \tau_{i}}}{\Delta \tau_{i}}\right) S_{i+1 / 2}
\end{eqnarray}

\subsection{{\bfc LVG criteria and integration limit}}
{In this work, we use the LVG approximation to predict the line profile of CO transitions. Our approximation is valid only when the velocity gradient in a characteristic length ($L$) is much greater than the local thermal velocity gradient within it as}
\begin{equation}
\Big| \frac{dV_{\rm n}}{dz} \Big| \gg v_{\rm th}/L
\end{equation}
{ where $L = T_{\rm n}/(dT_{\rm n}/dz)$. In Figure \ref{fig:lvg}, we compare the gradient of the gas velocity ($dV_{\rm n}/dz$) with that of the thermal velocity $V_{\rm th}/L$. One can see that the velocity gradient is sufficient such that LVG is satisfied only within a certain location $z$ in the shock. Beyond this distance, the LVG condition failed, and our scheme is invalid. Thus, we stop integrating Equation \ref{eq:ok87_o2} when the $\frac{dV_{\rm n}}{dz}< v_{\rm th}/L$.}

\begin{figure}[htp!]
\centering
{\includegraphics[width=0.45\textwidth]{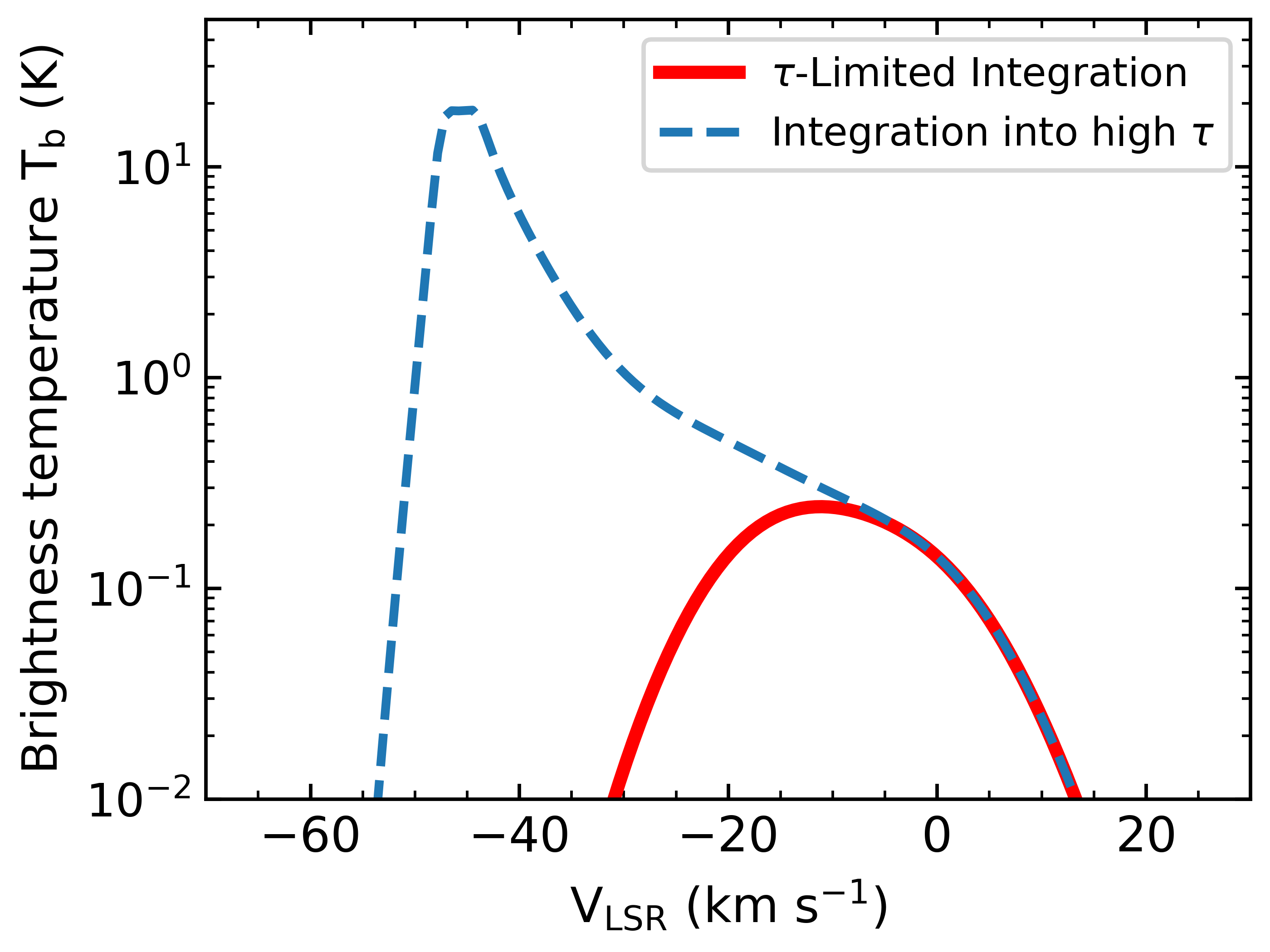}}
\caption{Effect of integration range for ($n$,$V_{\rm shock}$) = ($10^4$ cm$^{-3}$, 50 km s$^{-1}$) model. The solid line results from an integration up to $z \sim 10^{16}\,\rm cm$, while the dashed line is achieved for an entire shock integration, which shows a skewed and non-thermal profile due to error accumulation. \label{fig:varying_profile}}
\end{figure}

{As an example, Figure \ref{fig:varying_profile} compares the line profiles produced when integrating the entire shock region and plane-parallel slabs. Integration into the invalid LVG condition pronounces erroneous and non-thermal profiles. The generation of correct line profile prediction thus depends heavily on the consideration of a suitable integration range. In this case, the integration range is limited to roughly z $\sim 10^{16}\,$cm.}

\section{Result of Line Profile Fits}\label{sec:ResultLineProfile}
\begin{deluxetable}{clccccc}[htp!]
\tablecaption{Best-fit models for some CO observations in IC 443 and W 28 \label{tab:bestfit}}
\tablewidth{0pt}
\tablehead{
\colhead{} & \colhead{Line} & \colhead{$n({\rm H}_2)$} & \colhead{$V_{\rm Shock}$} & \colhead{b} & \colhead{V$_{\rm offset}$} & \colhead{$\theta$} \\
\nocolhead{} & \nocolhead{} & \colhead{\textit{cm$^{-3}$}} & \colhead{\textit{km s$^{-1}$}} & \nocolhead{($\mu$G cm$^{1.5}$)} & \nocolhead{km   s$^{-1}$} & \nocolhead{} 
}
\startdata
\multicolumn{7}{c}{\textbf{W 28}} \\ \cmidrule{1-7}
    \multirow{2}{*}{BML4} & CO(2-1)$\rm ^{1}$ & \multirow{2}{*}{10,000} & \multirow{2}{*}{$50$} & \multirow{2}{*}{$2.8$} & -3 & 190$^{\circ}$ \\ 
          & CO(16-15)$\rm ^{5}$ & & & & +3 & 0$^{\circ}$  \\ 
    \cmidrule{2-7}
    \multirow{2}{*}{OH(F)$\rm ^{2}$} & CO(3-2) & \multirow{2}{*}{10,000} & \multirow{2}{*}{$50$} & \multirow{2}{*}{$2.5$} & \multirow{2}{*}{+2} & \multirow{2}{*}{$60^{\circ}$} \\
    &  CO(11-10)  &  &  & & \\ 
    \cmidrule{1-7} 
\multicolumn{7}{c}{\textbf{IC 443}} \\ \cmidrule{1-7}
    \multirow{2}{*}{GI$\rm ^{3}$} & CO(2-1) & \multirow{2}{*}{8,000} & \multirow{2}{*}{$35$} & \multirow{2}{*}{$2.5$} & \multirow{2}{*}{-4} & \multirow{2}{*}{$30^{\circ}$} \\
    &  CO(3-2)  &  &  & & \\ 
    \cmidrule{2-7}
    H1$\rm ^{4}$ & CO(2-1) & \multirow{2}{*}{10,000} & \multirow{2}{*}{$35$} & \multirow{2}{*}{$2.5$} & \multirow{2}{*}{+3} & \multirow{2}{*}{$30^{\circ}$} \\
    H3$\rm ^{4}$ & CO(2-1) &  &  &  & \\ 
\enddata
\tablenotetext{}{1: \cite{reach_shocked_2005}, 2: \cite{gusdorf_probing_2012}, 3: \cite{van_dishoeck_submillimeter_1993}, 4: \cite{rho_near-infrared_2001}, 5: This work.}
\end{deluxetable}

\subsection{Line Profile Fits for W28 \label{subsec:w28_result}}

Generally, we found that previous estimates (by excitation diagrams and emission-line diagnostics) of shock properties for each observation agrees with the prediction from CO observations. The agreement between different post-processing methods - prediction of CO line profile and estimates from excitation diagram - shows strongly in preshock density and thus the magnetic field strength. 

Due to the 1D nature of the method, however, there is room for mismatch. It is possible to attribute a 2-component, 2-shock-velocity model using an excitation diagram to an observation that is a single-component with large shock velocity \citep{van_dishoeck_submillimeter_1993}. In our case, our model is only tuned to the molecular shock and we tried to find spec dominated by a single shock to test theoretical model, as our single shock model best reproduces the line-width of observation.

Absolute brightness depends on other detailed precision parameters (beam filling factors, etc.) which are unknown. We do not account for absolute brightness with this theoretical demonstration, only profile line-width as it directly constrains shock velocity. Thus maximum brightness intensity is set to 1. The integrated intensity of line profile (\(\int TdV\)) from each observation and each model and is included in Table \ref{tab:sofia_observations_properties}, \ref{tab:other_observation_properties} and Table \ref{tab:shockmodels}, which serves as a sanity check for choosing which model fits observation, on the basis that scale factors from 0.1 to 2 are reasonable.

\begin{figure*}[htp!]
\plotone{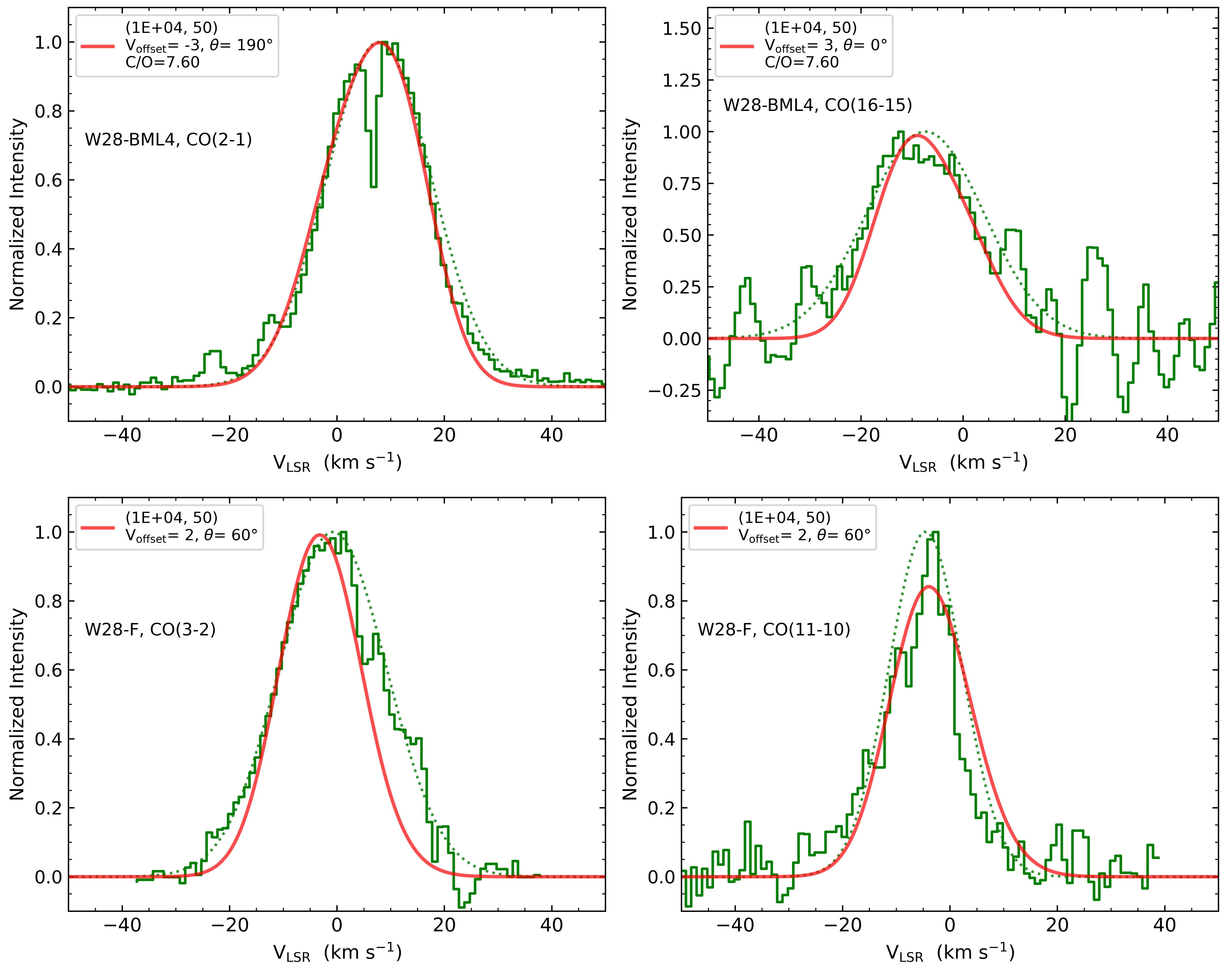}
\epsscale{1.2}
\caption{CO spectra in W28 (green), together with model fits (red). \textit{Left-to-right, top-to-bottom:} CO(2-1) and CO(16-15) lines observed with IRAM 30-m and SOFIA, respectively, toward W28-BML4; CO(3-2) and  CO(11-10) observed with APEX toward W28F. {The green dotted lines are simple Gaussian fits to the CO lines.}
\label{fig:w28profile}}
\end{figure*}
\begin{figure*}[ht!]
\plotone{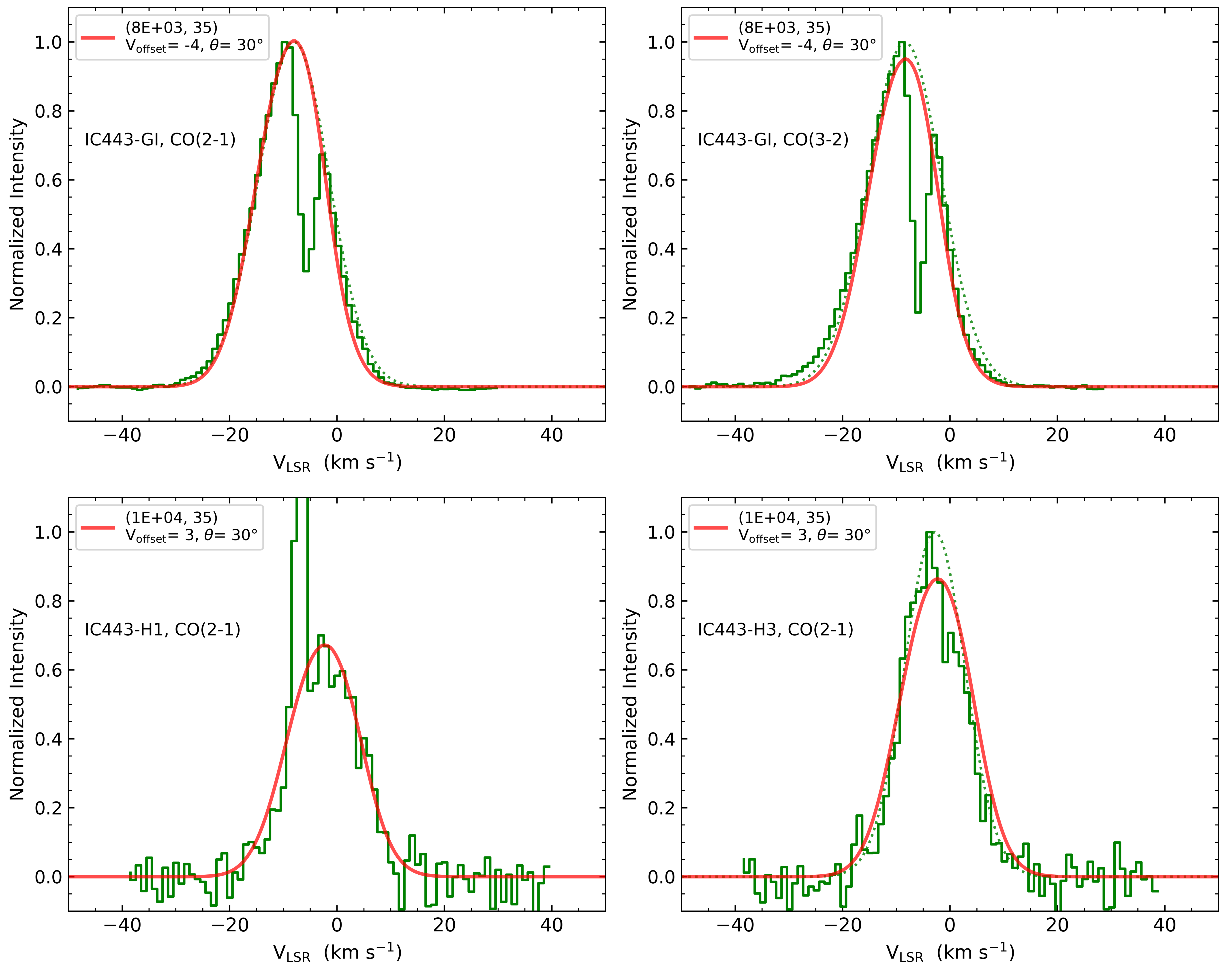}
\epsscale{1.2}
\caption{CO spectra in IC443 (green) and model fits (red). \textit{Left-to-right, top-to-bottom:} CO(2-1) and CO(3-2) observed with CSO towards clump and GI of IC433; CO(2-1) lines observed with the 12--m array on Kitt Peak (KP12/KPNO) towards clump H1 and H3 of IC443. \label{fig:ic443profile}}
\end{figure*}

From a priori shocks models in Table \ref{tab:shockmodels}, we produced line profile fits for observations summarized in Table \ref{tab:bestfit}. Further constraints from molecular lines ratio are used when available. 

For the high-J CO(16-15) observation of W28-BML4 (Figure \ref{fig:w28profile}), the linewidth limits shock velocity to V$_{\rm{Shock}} \sim$  50 km s$^{-1}$. A key constraint is the ratio $\langle \text{C}_{\rm II} \rangle/\langle  \text{O}_{\rm I} \rangle\sim 5$. 
Both the n(H$_{2}) \sim 10^4$ and $\num{5e3}$ cm$^{-3}$ models match the line-width, but from Table \ref{tab:shockmodels}, the latter produces a $\langle \text{C}_{\rm II} \rangle/\langle  \text{O}_{\rm I} \rangle$ ratio 10 times larger than observation. It is more likely that the emission originates from shock into denser gas at $10^4$ cm$^{-3}$. 

We found that the same model can reproduce CO(2-1) observation from \cite{reach_shocked_2005} in the BML4 region. The emission, though, appears to be reversed in the direction inferred from its central velocity. The absorption feature at +5 km s$^{-1}$ (FWHM = 1.2 km s$^{-1}$) infers the velocity of un-shocked, cold, ambient gas.

\subsection{Line Profile Fits for IC443}\label{subsec:ic443_result}

We follow the same roadmap for the prediction of the CO line profile with IC443. Excitation models were used to estimate shock properties in both observations of \cite{van_dishoeck_submillimeter_1993} and \cite{rho_shocked_2021}. \cite{reach_supernova_2019} has developed theoretical line profile prediction for H$_2$ from various observations of IC443 B, C, and G, which will serve as useful shock diagnostic. Line profile modeling of near-infrared emission H$_2$ S(5) with SOFIA of \cite{reach_supernova_2019} provided a constraints on preshock density $n(H_2)\approx 10^3-10^4 $ cm\(^{-3}\), and shock velocity in the range of 30--60 km/s, sufficient to produce the observed H$_2$ line-widths.

The CO observation of \cite{van_dishoeck_submillimeter_1993} provided exhaustive constraints on the origin of the CO emission in IC443. From $^{12}$CO and $^{13}$CO line ratios and excitation calculation for clump B and C, best fits of excitation calculation indicate preshock density n$(H_2)\sim 10^3$--$10^4$ at an initial gas temperature less than 20K. Quite importantly, preshock density of clump G cannot exceed $\num{5e4}$ cm\(^{-3}\) due to self-absorption of HCO$^{+}$ and HCN, indicating low excitation temperature (Section 5.2 of \cite{van_dishoeck_submillimeter_1993}). 

In this work, we found the linewidth from a-priori models sufficient to reproduce observations in clump GI. For both CO(2-1) and CO(3-2), one-component models n(H$_{2}) \sim \num{8e3}$ cm$^{-3}$ and V$_{\rm{Shock}} \sim 35$ km s$^{-1}$ make good fits with observation (see Figure \ref{fig:ic443profile}). This prediction is in agreement with prediction from H$_2$ theoretical modeling of \cite{reach_supernova_2019} ( $\num{6e3}$ cm$^{-3}$, 37 km s$^{-1}$) and infer that denser, high-frequency CS-producing components ($\sim 10^5$ cm$^{-3}$) are well-separated from low-J-CO-producing component in this region, a conclusion both reflected in \cite{turner_detection_1992} and \cite{van_dishoeck_submillimeter_1993}, where lower density gas are suggested to be responsible to low-frequency mission of CO (eg. 3-2 and 2-1), and higher density for high-frequency CS, (eg. 5-4 and 7-6). The strong self-absorption feature of both CO(2-1) and 3-2 in  clump GI, around -5 km s$^{-1}$, suggests the ambient gas velocity. 

Though, it should be noted that observations in other clumps of IC-443 from \cite{van_dishoeck_submillimeter_1993} are explained with a two-component model, with linewidth up to $\sim 80$ km s$^{-1}$ for which slow C-shock models alone cannot reproduce.

Shock fits of n(H$_{2}) \sim 10^4$ cm$^{-3}$ and V$_{\rm{Shock}} \sim $ 35 km s$^{-1}$ were made for CO(2-1) observation of the H3 position from \cite{rho_near-infrared_2001}. This prediction is in agreement with H$_2$ lines data and excitation models of \cite{rho_near-infrared_2001} and \cite{cesarsky_isocam_1999}, suggesting fast J-shock (n(H$_{2})  < 10^3$ cm$^{-3}$, V$_{\rm Shock}$ up to 100 km s$^{-1}$) and slow C-shock (n(H$_{2}) \sim 10^4$ cm$^{-3}$, V$_{\rm Shock}$ up to 50 km s$^{-1}$) \citep[][Figure 2]{mckee_interstellar_1991} through a dense medium with varying density, in which denser, not-dissociated CO gas mostly probe for slow C-shocks. 

\section{Conclusion\label{conclusion}}

Development of line profile prediction directly from shock models would strengthen many links in the studies of SNR-molecular clouds interaction. Shocks are ubiquitous major drivers of star formation, sources of cosmic ray acceleration, and sculptors of filamentary magnetic field structure. From the Paris-Durham magneto-hydrodynamic (MHD) 1D shock model, constraints such as preshock density, shock velocity, B-field strength and cosmic ionization rates are all available as shock diagnostics for related studies. 

We presented a post-processing framework to produce the theoretical line profiles of purely-rotational emission of CO from the Paris-Durham shock model. This is done by considering the Large Velocity Gradient approximation and the effect of optically-thick plane-parallel slabs in the integration range of the 1D model. 


{We found that rotational emissions of CO are mainly generated in the rapid cooling phase of shock. In comparison to H$_{2}$ emission lines, CO emits much later in the shock frame than H$_{2}$, due to its lower excitation energy.}

We tested the framework on CO(16-15), CO(11-10), CO(3-2) and CO(2-1) observations in regions BML4 and OH(F) of SNR W28 for shock velocity $50\,\rm km\,s^{-1}$ and preshock density $10^4\,$cm$^{-3}$, finding the prediction well-represented by shock studies of \cite{reach_shocked_2005} and \cite{gusdorf_probing_2012}.

We also modeled CO line profiles for CO(3-2) and CO(2-1) in clump GI and H of SNR IC443 with shock velocity $35\,\rm km\,s^{-1}$ and preshock density $\num{8e3}-10^4\,$cm$^{-3}$. The prediction are in close agreement with the excitation models from \cite{van_dishoeck_submillimeter_1993} and \cite{rho_near-infrared_2001}, and also in agreement with H$_2$ theoretical modeling from \cite{reach_supernova_2019}. 

{We showed that the CO line profile prediction provides robust constraints on the shock velocity and becomes a valuable additional tool for probing shock into molecular. Our proposed scheme can directly be extended to other molecular lines of interests, such as SiO.}

\begin{acknowledgements}
Based in part on observations made with the NASA/DLR Stratospheric Observatory for Infrared Astronomy (SOFIA). SOFIA is jointly operated by the Universities Space Research Association, Inc. (USRA), under NASA contract NNA17BF53C, and the Deutsches SOFIA Institut (DSI) under DLR contract 50 OK 0901 to the University of Stuttgart. 
Financial support for this work was provided by NASA (through award \#06-0001) issued by USRA. The author expresses gratitude for extensive virtual mentoring amidst the pandemic, without which this research would not have been completed.
\end{acknowledgements}

\bibliography{COShock}{}

\begin{thebibliography}{}
\expandafter\ifx\csname natexlab\endcsname\relax\def\natexlab#1{#1}\fi
\providecommand{\url}[1]{\href{#1}{#1}}
\providecommand{\dodoi}[1]{doi:~\href{http://doi.org/#1}{\nolinkurl{#1}}}
\providecommand{\doeprint}[1]{\href{http://ascl.net/#1}{\nolinkurl{http://ascl.net/#1}}}
\providecommand{\doarXiv}[1]{\href{https://arxiv.org/abs/#1}{\nolinkurl{https://arxiv.org/abs/#1}}}

\bibitem[{Brogan {et~al.}(2004)Brogan, Goss, Lazendic, \&
  Green}]{brogan_oh_2004}
Brogan, C.~L., Goss, W.~M., Lazendic, J.~S., \& Green, A.~J. 2004, The
  Astronomical Journal, 128, 700

\bibitem[{Burton(1987)}]{burton_ic_1987}
Burton, M. 1987, Quarterly Journal of the Royal Astronomical Society, 28, 269

\bibitem[{Cesarsky {et~al.}(1999)Cesarsky, Cox, Pineau~des Forêts, van
  Dishoeck, Boulanger, \& Wright}]{cesarsky_isocam_1999}
Cesarsky, D., Cox, P., Pineau~des Forêts, G., {et~al.} 1999, Astronomy \&
  Astrophysics, 348, 945

\bibitem[{Chevalier(1999)}]{chevalier_supernova_1999}
Chevalier, R.~A. 1999, Astrophysical Journal, 511, 798

\bibitem[{Claussen {et~al.}(1997)Claussen, Frail, Goss, \&
  Gaume}]{claussen_polarization_1997}
Claussen, M.~J., Frail, D.~A., Goss, W.~M., \& Gaume, R.~A. 1997, Astrophysical
  Journal, 489, 143

\bibitem[{Draine \& McKee(1993)}]{draine_theory_1993}
Draine, B.~T., \& McKee, C.~F. 1993, Annual Review of Astronomy and
  Astrophysics, 31, 373

\bibitem[{Draine {et~al.}(1983)Draine, Roberge, \&
  Dalgarno}]{draine_magnetohydrodynamic_1983}
Draine, B.~T., Roberge, W.~G., \& Dalgarno, A. 1983, Astrophysical Journal,
  264, 485

\bibitem[{Drury \& Strong(2017)}]{drury_power_2017}
Drury, L.~O., \& Strong, A.~W. 2017, Astronomy \& Astrophysics, 597, A117

\bibitem[{Dubner {et~al.}(2000)Dubner, Velázquez, Goss, \&
  Holdaway}]{dubner_high-resolution_2000}
Dubner, G.~M., Velázquez, P.~F., Goss, W.~M., \& Holdaway, M.~A. 2000, The
  Astronomical Journal, 120, 1933

\bibitem[{Flower \& Forêts(2015)}]{flower_interpreting_2015}
Flower, D.~R., \& Forêts, G. P.~d. 2015, Astronomy \& Astrophysics, 578, A63

\bibitem[{Giuliani {et~al.}(2010)Giuliani, Tavani, Bulgarelli, Striani,
  Sabatini, Cardillo, Fukui, Kawamura, Ohama, Furukawa, Torii, Sano, Aharonian,
  Verrecchia, Argan, Barbiellini, Caraveo, Cattaneo, Chen, Cocco, Costa,
  D'Ammando, Del~Monte, de~Paris, Di~Cocco, Donnarumma, Evangelista, Feroci,
  Fiorini, Froysland, Fuschino, Galli, Gianotti, Labanti, Lapshov, Lazzarotto,
  Lipari, Longo, Marisaldi, Mereghetti, Morselli, Moretti, Pacciani,
  Pellizzoni, Perotti, Picozza, Pilia, Prest, Pucella, Rapisarda, Rappoldi,
  Soffitta, Trifoglio, Trois, Vallazza, Vercellone, Vittorini, Zambra, Zanello,
  Pittori, Santolamazza, Giommi, Colafrancesco, \&
  Salotti}]{giuliani_agile_2010}
Giuliani, A., Tavani, M., Bulgarelli, A., {et~al.} 2010, Astronomy \&
  Astrophysics, 516, L11

\bibitem[{Godard {et~al.}(2019)Godard, Pineau~des Forêts, Lesaffre, Lehmann,
  Gusdorf, \& Falgarone}]{godard_models_2019}
Godard, B., Pineau~des Forêts, G., Lesaffre, P., {et~al.} 2019, Astronomy \&
  Astrophysics, 622, A100

\bibitem[{Gusdorf {et~al.}(2012)Gusdorf, Anderl, Güsten, Stutzki, Hübers,
  Hartogh, Heyminck, \& Okada}]{gusdorf_probing_2012}
Gusdorf, A., Anderl, S., Güsten, R., {et~al.} 2012, Astronomy \& Astrophysics,
  542, L19

\bibitem[{Gusdorf {et~al.}(2008)Gusdorf, Cabrit, Flower, \& Pineau
  Des~Forêts}]{gusdorf_sio_2008}
Gusdorf, A., Cabrit, S., Flower, D.~R., \& Pineau Des~Forêts, G. 2008,
  Astronomy \& Astrophysics, 482, 809

\bibitem[{Hummer \& Rybicki(1985)}]{hummer_sobolev_1985}
Hummer, D.~G., \& Rybicki, G.~B. 1985, Astrophysical Journal, 293, 258

\bibitem[{Lesaffre {et~al.}(2013)Lesaffre, Pineau~des Forêts, Godard,
  Guillard, Boulanger, \& Falgarone}]{lesaffre_low-velocity_2013}
Lesaffre, P., Pineau~des Forêts, G., Godard, B., {et~al.} 2013, Astronomy \&
  Astrophysics, 550, A106

\bibitem[{Mckee \& Draine(1991)}]{mckee_interstellar_1991}
Mckee, C.~F., \& Draine, B.~T. 1991, Science, 252, 397

\bibitem[{Neufeld \& Kaufman(1993)}]{neufeld_radiative_1993}
Neufeld, D.~A., \& Kaufman, M.~J. 1993, Astrophysical Journal, 418, 263

\bibitem[{Olson \& Kunasz(1987)}]{olson_short_1987}
Olson, G.~L., \& Kunasz, P.~B. 1987, Journal of Quantitiative Spectroscopy and
  Radiative Transfer, 38, 325

\bibitem[{Pannuti {et~al.}(2017)Pannuti, Rho, Kargaltsev, Rangelov, Kosakowski,
  Winkler, Keohane, Hare, \& Ernst}]{pannuti_ctio_2017}
Pannuti, T.~G., Rho, J., Kargaltsev, O., {et~al.} 2017, Astrophysical Journal,
  839, 59

\bibitem[{Phan {et~al.}(2020)Phan, Gabici, Morlino, Terrier, Vink, Krause, \&
  Menu}]{phan_constraining_2020}
Phan, V. H.~M., Gabici, S., Morlino, G., {et~al.} 2020, Astronomy \&
  Astrophysics, 635, A40

\bibitem[{Reach \& Rho(1996)}]{reach_shockingly_1996}
Reach, W.~T., \& Rho, J. 1996, Astronomy \& Astrophysics, 315, L277

\bibitem[{Reach {et~al.}(2005)Reach, Rho, \& Jarrett}]{reach_shocked_2005}
Reach, W.~T., Rho, J., \& Jarrett, T.~H. 2005, Astrophysical Journal, 618, 297

\bibitem[{Reach {et~al.}(2019)Reach, Tram, Richter, Gusdorf, \&
  DeWitt}]{reach_supernova_2019}
Reach, W.~T., Tram, L.~N., Richter, M., Gusdorf, A., \& DeWitt, C. 2019,
  Astrophysical Journal, 884, 81

\bibitem[{Rho {et~al.}(2001)Rho, Jarrett, Cutri, \&
  Reach}]{rho_near-infrared_2001}
Rho, J., Jarrett, T.~H., Cutri, R.~M., \& Reach, W.~T. 2001, Astrophysical
  Journal, 547, 885

\bibitem[{Rho {et~al.}(2021)Rho, Jarrett, Tram, Lim, Reach, Bieging, Lee, Koo,
  \& Whitney}]{rho_shocked_2021}
Rho, J., Jarrett, T.~H., Tram, L.~N., {et~al.} 2021, arXiv e-prints, 2105,
  arXiv:2105.10617

\bibitem[{Rho {et~al.}(1996)Rho, Petre, Pisarski, \&
  Jones}]{rho_ROSATObservationW28_1996}
Rho, J.~H., Petre, R., Pisarski, R., \& Jones, L.~R. 1996, {{ROSAT}}
  Observation of {{W28}}., 273--274

\bibitem[{Rowell {et~al.}(2000)Rowell, Naito, Dazeley, Edwards, Gunji, Hara,
  Holder, Kawachi, Kifune, Matsubara, Mizumoto, Mori, Muraishi, Muraki,
  Nishijima, Ogio, Patterson, Roberts, Sako, Sakurazawa, Susukita, Tamura,
  Tanimori, Thornton, Yanagita, Yoshida, \&
  Yoshikoshi}]{rowell_ObservationsSupernovaRemnant_2000}
Rowell, G.~P., Naito, T., Dazeley, S.~A., {et~al.} 2000, Astronomy and
  Astrophysics, 359, 337

\bibitem[{Sobolev(1957)}]{sobolev_diffusion_1957}
Sobolev, V.~V. 1957, Soviet Astronomy, 1, 678

\bibitem[{Surdej(1977)}]{surdej_contribution_1977}
Surdej, J. 1977, Astronomy \& Astrophysics, 60, 303

\bibitem[{Temim {et~al.}(2019)Temim, Slane, Sukhbold, Koo, Raymond, \&
  Gelfand}]{temim_ProbingInnermostEjecta_2019}
Temim, T., Slane, P., Sukhbold, T., {et~al.} 2019, The Astrophysical Journal,
  878, L19

\bibitem[{Tram {et~al.}(2018)Tram, Lesaffre, Cabrit, Gusdorf, \&
  Nhung}]{tram_h2_2018}
Tram, L.~N., Lesaffre, P., Cabrit, S., Gusdorf, A., \& Nhung, P.~T. 2018,
  Monthly Notices of the Royal Astronomical Society, 473, 1472

\bibitem[{Troja {et~al.}(2008)Troja, Bocchino, Miceli, \&
  Reale}]{troja_xmm-newton_2008}
Troja, E., Bocchino, F., Miceli, M., \& Reale, F. 2008, Astronomy \&
  Astrophysics, 485, 777

\bibitem[{Turner(1992)}]{turner_detection_1992}
Turner, B.~E. 1992, Astrophysical Journal, 396, L107

\bibitem[{Ustamujic {et~al.}(2020)Ustamujic, Orlando, Greco, Miceli, Bocchino,
  Tutone, \& Peres}]{ustamujic_modeling_2020}
Ustamujic, S., Orlando, S., Greco, E., {et~al.} 2020, arXiv e-prints, 2012,
  arXiv:2012.08017

\bibitem[{van~der Tak {et~al.}(2007)van~der Tak, Black, Schöier, Jansen, \&
  van Dishoeck}]{van_der_tak_computer_2007}
van~der Tak, F. F.~S., Black, J.~H., Schöier, F.~L., Jansen, D.~J., \& van
  Dishoeck, E.~F. 2007, Astronomy \& Astrophysics, 468, 627

\bibitem[{van Dishoeck {et~al.}(1993)van Dishoeck, Jansen, \&
  Phillips}]{van_dishoeck_submillimeter_1993}
van Dishoeck, E.~F., Jansen, D.~J., \& Phillips, T.~G. 1993, Astronomy \&
  Astrophysics, 279, 541

\bibitem[{Vaupré {et~al.}(2014)Vaupré, Hily-Blant, Ceccarelli, Dubus, Gabici,
  \& Montmerle}]{vaupre_cosmic_2014}
Vaupré, S., Hily-Blant, P., Ceccarelli, C., {et~al.} 2014, Astronomy \&
  Astrophysics, 568, A50

\bibitem[{Wardle \& {Yusef-Zadeh}(2002)}]{wardle_SupernovaRemnantOH_2002}
Wardle, M., \& {Yusef-Zadeh}, F. 2002, Science, 296, 2350

\bibitem[{Zhou {et~al.}(2014)Zhou, Safi-Harb, Chen, Zhang, Jiang, \&
  Ferrand}]{zhou_xmm-newton_2014}
Zhou, P., Safi-Harb, S., Chen, Y., {et~al.} 2014, Astrophysical Journal, 791,
  87

\end{thebibliography}
\bibliographystyle{aasjournal}
\end{document}